# Generalized Batchelor functions of isotropic turbulence

Elias Gravanis[1] and Evangelos Akylas[1]

[1]*Department of Civil Engineering and Geomatics,*
*Cyprus University of Technology*
*P.O. Box 50329, 3603, Limassol, Cyprus*
e-mail: elias.gravanis@cut.ac.cy;  evangelos.akylas@cut.ac.cy

**ABSTRACT**

We generalize Batchelor's parameterization of the autocorrelation functions of isotropic turbulence in a form involving a product expansion with multiple small scales. The richer small scale structure acquired this way, compared to the usual Batchelor function, is necessary so that the associated energy spectrum approximate well actual spectra in the universal equilibrium range. We propose that the generalized function provides an approximation of arbitrary accuracy for actual spectra of isotropic turbulence over the universal equilibrium range. The degree of accuracy depends on the number of higher moments which are determinable and it is reflected in the number of small scales involved. The energy spectrum of the generalized function is derived, and for the case of two small scales is compared with data from high-resolution direct numerical simulations. We show that the compensated spectra (which illustrate the bottleneck effect) and dissipation spectra are encapsulated excellently, in accordance with our proposal.

## I. INTRODUCTION

Decades ago, Batchelor [1] wrote down a simple formula for the second order structure function of the velocity field in isotropic turbulence. The formula incorporates the power series nature of the second order correlation functions in the small separations on the one hand and the Kolmogorov 2/3 law [2] in the inertial range on the other. The basic idea can be applied equally well to the longitudinal, transversal or three dimensional structure functions. In the relatively recent past, the spectrum associated with the longitudinal Batchelor structure function was calculated analytically [3][4]. The idea of using Batchelor interpolation has been applied, including also anomalous scaling, to the longitudinal structure functions, see e.g. [5-9], mostly in association with the residual dependence of various quantities characterizing turbulence on the Reynolds number for large but finite values; also, to transversal structure functions, see e.g. [10-12], in association with the bottleneck effect, illustrated by the characteristic bump in the graphs of the compensated spectrum [13], which is a standard feature of the Batchelor type of spectra.

The Batchelor function is entirely fixed by the total turbulent kinetic energy, the dissipation rate, the viscosity and the Kolmogorov constant $C_2$ arising in the two-thirds law. This means that the higher moments of the energy spectrum, such as the palinstrophy, are fixed in terms of these quantities. Dimensionless quantities depend solely on $C_2$. That is, dimensionless characteristic numbers, such as the velocity derivative skewness or the position of the bottleneck bump peak in dimensionless wavenumbers, are entirely fixed in terms of $C_2$. As we shall explicitly see below this is not consistent with the behavior of isotropic turbulent flows. Therefore those pre-fixed higher moments restrict the applicability of Batchelor functions; for example, its energy spectrum cannot accommodate the characteristics of the bottleneck bump. These difficulties can be naturally and easily resolved by a simple generalization along the lines of construction of the original Batchelor function. This generalization is the subject of this work.

## II. BATCHELOR FUNCTIONS AND SPECTRA

We start by introducing the original Batchelor function and deriving the associated energy spectrum. The longitudinal second order structure function is $\overline{(u_l(\mathbf{r}_1)-u_l(\mathbf{r}_2))^2}$, where $u_l$ is the component of the velocity field in the direction of separation and overbar denotes a suitable average. The longitudinal (normalized) autocorrelation function $f$ is defined by $\overline{u_l(\mathbf{r}_1)u_l(\mathbf{r}_2)}=u^2 f(r)$ where $r$ is the distance between the separation points and $u^2$ is the average of $u_l^2$ i.e. the mean value of the square of the velocity in any specific direction. The dissipation rate $\varepsilon$ can be expressed as $\varepsilon = -15\nu u^2 f''(0)$ where $\nu$ is the viscosity.

Using symmetry considerations it is easy to show that second order structure function and $f$ are even functions of $r$ and can be expressed as a power series around the origin $r = 0$; also the series contain alternating sign terms of $r^2$. Adopting the Kolmogorov scaling we have that $\overline{(u_l(\mathbf{r}_1)-u_l(\mathbf{r}_2))^2} = C_2(\varepsilon r)^{2/3}$ when $r$ is in the inertial range. That is, these functions become non-analytic at such distances and therefore they must have a finite radius of convergence when expressed as power series.

Guessing a form for $f$ based on these conditions, a rather minimal choice is the Batchelor function [1] which we may write as

$$f = 1 - \frac{\varepsilon}{30\nu u^2} r^2 \left(1 + \frac{r^2}{\theta^2}\right)^{-2/3} \tag{1}$$

We shall work with the longitudinal functions throughout. $\theta$ is the power series radius of convergence mentioned above. The inertial range scaling arises at the infinity of the coordinate $r/\theta$. This is already telling us that $\theta$ should probably be a dissipation range scale. The coefficient involving $\varepsilon$ and $\nu$ is justified below. By its very construction, the function (1) claims validity in the entire universal equilibrium range but not outside of it.

The three-dimensional energy spectrum is given by (see e.g. [14])

$$E(k) = \frac{u^2}{\pi} \int_0^\infty f(r)(kr \sin kr - k^2 r^2 \cos kr) dr \tag{2}$$

The function (1) contains no information about the large scales outside the inertial range. Hence, it does not claim carrying information about the smallest wave-numbers. The integral (2) diverges in the large distances. We must then isolate those spurious 'infrared' divergences and throw them away. The first term in $f$ i.e., the constant, introduces a Dirac delta function at $k = 0$. That can be thrown away. The rest can be taken care of by regularizing the integral through analytic continuation.

Using the definition of the Euler Gamma function we may write

$$r^2\left(1+\frac{r^2}{\theta^2}\right)^{-\zeta} = r^2 \int_0^\infty dt \frac{t^{\zeta-1}}{\Gamma(\zeta)} e^{-t} e^{-r^2 t/\theta^2} = -\theta^2 \int_0^\infty dt \frac{t^{\zeta-1}}{\Gamma(\zeta)} e^{-t} \frac{d}{dt} e^{-r^2 t/\theta^2} \tag{3}$$

where we have used a general exponent $\zeta$ in the place of the 2/3. The Batchelor function is essentially a continuous superposition of Gaussian 'correlation functions'. Therefore the spectrum of the Batchelor function is a superposition of the spectra of those functions. One finds easily that

$$E(k) = \frac{\varepsilon \theta^7 k^4}{240\sqrt{\pi}\Gamma(\zeta)\nu} \int_0^\infty dt\, t^{\zeta-1} e^{-t} \frac{d}{dt}\{t^{-5/2} e^{-\frac{k^2\theta^2}{4t}}\} \tag{4}$$

Setting now $\zeta = 2/3$ and using standard integral expressions of the modified Bessel functions $K_\alpha(x)$ one finds

$$E(k) = \frac{1}{30 \times 2^{\frac{1}{6}}\sqrt{\pi}\Gamma(\frac{2}{3})} \frac{\varepsilon}{\nu} \theta^3 (k\theta)^{\frac{7}{6}} \left\{k\theta K_{\frac{11}{6}}(k\theta) + \tfrac{2}{3} K_{\frac{17}{6}}(k\theta)\right\} \tag{5}$$

This expression is equivalent to the one found in [3]. In the limit $k\theta \to 0$ emerges the Kolmogorov 5/3 law and the length $\theta$ is fixed in terms of $C_2$ and the Kolmogorov dissipation scale $\eta = (\nu^3/\varepsilon)^{1/4}$:

$$E(k) = C_K \varepsilon^{2/3} k^{-5/3}, \qquad C_K = \frac{55}{27\Gamma(\frac{1}{3})} C_2, \qquad \theta = (15C_2)^{\frac{3}{4}} \eta \tag{6}$$

The (exact) relation between the standard constants $C_K$ and $C_2$ which emerges here is well known [14].

The general expansion of the autocorrelation function $f$ around $r = 0$ up to the fourth order reads

$$f = 1 - \frac{\varepsilon}{30\nu u^2} r^2 + \frac{P}{420 u^2} r^4 + ..., \qquad P = \int_0^\infty k^4 E(k) dk \qquad (7)$$

where $P$ is the palinstrophy, the first higher moment of the spectrum. Expanding the function (1) one finds

$$P = \frac{28\varepsilon}{3\nu\theta^2} \qquad (8)$$

That is, the palinstrophy is fixed in terms of $\varepsilon$, $\nu$ and $C_2$. The same applies to all higher moments; in fact, their dimensionless forms are simply fixed by $C_2$. Clearly this is too restrictive. For example, the value of $P$ predicted by (8) is rather small and the derived spectrum has wrong behavior, as we shall explicitly discuss below. The Batchelor functions should be generalized in order to acquire more structure in the small scales.

Consider now the function

$$f = 1 - \frac{\varepsilon}{30\nu u^2} r^2 \prod_{i=1}^{N} \left(1 + \frac{r^2}{\theta_i^2}\right)^{-\zeta_i} \qquad (9)$$

for $N$ different lengths $\theta_i$ and exponents $\zeta_i$. That is, the autocorrelation function is constructed via a product expansion. The good thing with a product expansion is that adding more factors increases the degree of approximation same as adding terms to a polynomial approximation, while simultaneously the negative exponents allow the function to 'see' much further than a polynomial approximation. Clearly, the number of factors are counted by the number of lengths $\theta$, and we shall mostly use the number of $\theta$'s to designate the order $N$ of the product in (9). For positive $\zeta_i$ the function (9) possesses the same alternating sign structure of expression (1) as a power series around $r = 0$, which is necessary requirement (on both) in order to be consistent with the general expansion (7).

From a mathematical point of view, the $\theta$'s and the $\zeta$'s are the locations and the weights of the singularities (9) in the complex $r$-plane. From a physical point of view, the introduction of multiple dissipative scales $\theta_i$ reminds one of the fluctuating dissipative scale in the multi-fractal approaches to turbulence [15] or the cut-off dependent dissipative scale inherent in the renormalization group approach to turbulence [16]. Presumably, as we shall see, the number $N$ of the factors in (9) is related to the resolution level of the direct numerical simulations of turbulence whose data one attempts to encode in (9) i.e. reflects a sort of a cut-off. The common theme of the mentioned general approaches to turbulence is that there is actually not a single dissipation scale. The scales $\theta_i$, along with the $\zeta_i$ which control the weight of the $\theta$'s in the product expansion of (9), essentially express that characteristic of turbulence. This is an intuitive way to think about the scales $\theta_i$. More directly, they can be related to the moments of the energy spectrum, through generalizations of equation (8). Also, the spectra of the moments possess characteristics, such the position of their peaks, to which the $\theta$'s may also be associated, although in a less direct mathematical manner. Throughout this work we shall think of the scales $\theta$ primarily as realizing information about the energy spectrum moments.

The function (9) reproduces the 2/3 law if $\zeta_1+\ldots+\zeta_N = 2/3$ and

$$\theta_1^{2\zeta_1}\theta_2^{2\zeta_2}\cdots\theta_N^{2\zeta_N} = 15C_2\eta^{\frac{4}{3}} \tag{10}$$

which is a generalization of the last relation in (6). Clearly, if all $\theta_i$ are equal then the function (9) degenerates to the original Batchelor function (1).

In practice, the energy spectrum associated with (9) can be calculated numerically directly from the formula (2). One needs only to multiply the expression (9) with a regularizing factor which vanishes at infinity, via some large length scale. If the scale is chosen to be large enough then the spectrum is left practically unaffected for all wavenumbers of interest. On the other hand, we may give an expression for the spectrum. The following identity is useful (see e.g. [17])

$$\prod_i Q_i^{-\zeta_i} = \int dV_t\,(t_1Q_1+\cdots t_NQ_N)^{-\sum_i \zeta_i}, \quad dV_t = \frac{\Gamma(\sum_i \zeta_i)}{\Gamma(\zeta_1)\cdots\Gamma(\zeta_N)}d^{N-1}t\,t_1^{\zeta_1-1}\cdots t_N^{\zeta_N-1} \tag{11}$$

The domain of integration is the $(N-1)$-simplex defined by $t_1+\cdots+t_N = 1$ and $t_i \geq 0$. The measure of integration $dV_t$ is normalized so that $\int dV_t = 1$. By this identity the product part of the function (9) reads

$$\int dV_t \left(1+\frac{r^2}{\Theta(t)^2}\right)^{-\sum_i \zeta_i} \tag{12}$$

where $\Theta(t)^{-2} = t_1\theta_1^{-2}+\cdots+t_N\theta_N^{-2}$. Formula (12) tells us that the non-trivial part of the generalized Batchelor function (9) is a superposition of the usual Batchelor functions, with a suitably defined length $\Theta(t)$. That in turn means that the spectrum of the generalized Batchelor function is a superposition of the spectra of the usual Batchelor function (setting at this point $\zeta_1+\cdots+\zeta_N = 2/3$):

$$E(k) = \frac{1}{30\times 2^{\frac{1}{6}}\sqrt{\pi}\Gamma(\frac{2}{3})}\frac{\varepsilon}{\nu}\int dV_t\,\Theta^3(k\Theta)^{\frac{7}{6}}\left\{k\Theta K_{\frac{11}{6}}(k\Theta)+\tfrac{2}{3}K_{\frac{17}{6}}(k\Theta)\right\} \tag{13}$$

In the limit of small wave-numbers this spectrum takes the form (6) via the identity (11) and the condition (10) as it should.

We may digress at this point to note the following. The measure $dV_t$ in equation (11) defines a Dirichlet distribution (multivariate version of the beta distribution) over the simplex of the interpolation parameters $t_i$. Equation (12) may then be interpreted as the expectation value of a (usual) Batchelor function with a $t$-dependent scale $\Theta(t)$ i.e., a continuous range of scales, which is defined through an interpolation between $N$ scales $\theta$ lying at the edges of the simplex. Equation (13) for the spectrum may be interpreted in a similar way as the expectation value of a usual Batchelor spectrum. There is a certain affinity between these expressions and models of the velocity increment fluctuations in the multi-fractal approach (see e.g. [7][9][18]), due to the nature of the dissipation scale as a continuous variable over a continuous range, and the presence of a probability distribution for that variable scale. The difference is that the multi-fractal models are formed on the basis of the intermittency phenomenon and the associated anomalous scaling, while in the present case anomalous scaling is only a possibility.

In the limit of the large wave-numbers, the modified Bessel functions approach an exponential function. Let us denote the smallest and largest of the lengths $\theta_i$ by $\theta_{min}$ and $\theta_{max}$ respectively. Then $\theta_{min} \leq \Theta(t) \leq \theta_{max}$. Then it is not difficult to show (using $\int dV_t = 1$) that the large $k$ spectrum is bounded by

$$\frac{1}{30 \cdot 2^{\frac{2}{3}} \Gamma(\frac{2}{3})} \frac{\varepsilon}{\nu} \theta_{min}^{\frac{14}{3}} k^{\frac{5}{3}} e^{-k\theta_{max}} \leq E(k) \leq \frac{1}{30 \cdot 2^{\frac{2}{3}} \Gamma(\frac{2}{3})} \frac{\varepsilon}{\nu} \theta_{max}^{\frac{14}{3}} k^{\frac{5}{3}} e^{-k\theta_{min}} \qquad (14)$$

Equality is attained in the case of the single length $\theta$ i.e. in the case of the usual Batchelor function. Thus we obtain that the generalized Batchelor function spectrum maintains, as a matter of magnitude, the (exponential)×(power law) asymptotics of the usual Batchelor spectrum (5). The estimate (14) makes sense as long as $k\theta_{min}$ is large. This estimate would require refinement, and most possibly will be modified qualitatively, in the case of an infinite number of lengths $\theta_i$ such that they approach zero; the expressions (9) and (14) may very well make sense even in the case of an infinite number of lengths $\theta$. For a finite number of $\theta$ one may verify numerically that the dominant behavior of the spectrum is $\exp(-k\theta_{min})$ – modulo power law factor corrections – which is rather expected intuitively.

We may now explain our viewpoint, or better our conjecture, in regard to the generalized Batchelor function (9) based on the product expansion. By adopting an adequately large product in (9), which may even be infinite, we may approximate to an ever increasing degree of approximation any autocorrelation function of isotropic turbulence, or more precisely, any energy spectrum of isotropic turbulence in the universal range, and especially in the largest meaningful wavenumbers. That is, the increasing degree of approximation rests on the increasing number of the scales $\theta$ (and the associated exponent $\zeta$). As the number of these scales increases, a greater number of higher moments of the spectrum (13) will agree with those of the actual spectrum one approximates.

Indeed, in this work we shall realize the last observation by taking into account the value of palinstrophy $P$ of the spectrum, given in equation (7), which is the first higher moment of the spectrum. It is straightforward to show that the dimensionless palinstrophy associated with (9) reads

$$\frac{\eta^5 P}{(\nu^5 \varepsilon)^{1/4}} = 14 \sum_{i=1}^{N} \zeta_i \frac{\eta^2}{\theta_i^2} \qquad (15)$$

Also the condition (10) constrains the dimensionless lengths $\theta_i/\eta$ in terms of $C_2$. Therefore if we work with a two-factor product in the function (9) i.e., $N=2$, knowing $\theta_i/\eta$ (and the exponents $\zeta_i$) is equivalent to knowing $C_2$ and the dimensionless palinstrophy. Similar formulas to (15) can be derived for all higher moments. We shall not need them in the present work.

Finally, it is worth to note that the generalized Batchelor spectrum (13), in its dimensionless form $(\nu^5\varepsilon)^{-1/4}E(k)$, depends solely on the quantities $\theta_i/\eta$ and $k\eta$. Therefore, the dimensionless Batchelor spectrum as a function of $k\eta$ depends solely on the value of $C_2$ and the values of a number of higher moments in dimensionless form in any given Reynolds number. That is, Reynolds-dependence enters the Batchelor spectrum (13) only through the Reynolds-dependence of these quantities.

## III. APPLICATIONS

We apply these ideas to the results of the high-resolution direct numerical simulations (DNS) of turbulence [19][20][21]. In that set up, turbulence reaches a stationary state by being fed energy at the largest scales. When Reynolds number is not too small there is a regime of scales where turbulence can be regarded as isotropic.

Under these conditions the following relation may be derived

$$|S| = \frac{12\sqrt{15}}{7} \frac{\eta^5 P}{(\nu^5 \varepsilon)^{1/4}} \qquad (16)$$

(see e.g. [5]) where is $S$ the skewness of the velocity derivative distribution. This number is a highly important descriptive parameter of turbulence as, by (16), is related to information from the dissipation sub-range: for the cases we shall consider the palinstrophy spectrum $k^4 E(k)$ peaks at $k\eta \approx 0.5$ i.e., depends on information from the small wavenumber end of the dissipation sub-range. Moreover, the shape, the peak and the position of the bottleneck bump [13] in the graph of the compensated spectrum $k^{5/3}E(k)$ depends mainly on the interplay between the two numbers $C_2$ (or $C_K$) and $S$, at the given Reynolds number. This statement will be explicitly realized below.

Combining the equations (15) and (16) one finds

$$|S| = 24\sqrt{15} \sum_{i=1}^{N} \zeta_i \frac{\eta^2}{\theta_i^2} \qquad (17)$$

Let us first consider the case of a single scale $\theta$ i.e. the usual Batchelor function. Combining equation (17) with the last relation in (6) – and taking the single exponent $\zeta$ to be $2/3$ – one finds $|S|=(16/15)C_2^{-3/2}$. Taking $C_2=2$ we find $|S|=0.38$. This is too small. The data of [20][21] find $|S|$ roughly in the range $0.52 - 0.6$. A misshaped bottleneck bump is rather expected, according to our arguments above. Even worse, the result is too 'rigid'. All higher moments are fixed in terms of $C_2$. Now, the value of the $C_2$ depends on the realization of the inertial sub-range at the given Reynolds number. Therefore $C_2$ should not determine parameters which carry information about the dissipation sub-range; certainly, whatever is determined will be likely fixed at a wrong value. Hence the energy spectrum will be incorrect deeper in the dissipation sub-range.

Going to two $\theta$'s we have more flexibility. We can give palinstrophy and skewness a much more acceptable value. The compensated spectrum $k^{5/3}E(k)$ as well as the dissipation spectrum $k^2E(k)$ can be now approximated well, as we shall see explicitly below. Palinstrophy and skewness will still be somewhat off their correct values, because with two $\theta$'s the palinstrophy *spectrum* $k^4E(k)$ cannot be approximated well, that is, it will have a wrong shape and therefore wrong – although much less wrong – area under it. The reason for that, is the fact that moments higher than palinstrophy will still be predicted in terms of the $C_2$ and $S$ at the given Reynolds number (at some incorrect values). Now, if the palinstrophy spectrum could be approximated well, then it turns out that the value of the next higher moment could not be as wrong as the predicted value (one may verify that by experimenting with simple model spectra). Therefore the palinstrophy spectrum $k^4E(k)$ cannot be approximated well, drawing also the value of the palinstrophy and skewness somewhat away from their correct values. (One should bear in mind, in the case of DNS spectra, 'correct values' for palinstrophy and skewness implies that there is enough resolution such that the errors are small. This means $k_{max}\eta$ at least 2). Improving that requires to go to three $\theta$'s. Then the problem is transferred to the next higher moments. Proceeding this way one may restrict the inaccuracies of the Batchelor spectrum deeper in the dissipation range, as far as the available data allow.

The DNS spectra presented in reference [21] include Reynolds numbers in the neighborhood of a thousand. These spectra have been obtained with resolution $k_{max}\eta \approx 1$, which is enough for our purposes. We will determine the generalized Batchelor spectrum which approximates these DNS spectra for the case of two $\theta$'s. The free parameters of this function are the lengths $\theta_1$, $\theta_2$ and the associated exponents $\zeta_1$, $\zeta_2$. The exponents $\zeta$ are assumed to be constrained by the 2/3 law, $\zeta_1+\zeta_2=2/3$. One may note though that this is not really necessary: by equation (5), the usual Batchelor function spectrum (6) and therefore the generalized Batchelor function spectrum (13) can be very easily written for a general (anomalous) exponent in the inertial range.

Determining the free parameters entails a best fit procedure. To do this systematically and effectively one should set up a best algorithm either at the level of the spectrum, which is more straightforward but cumbersome, either at the level of the autocorrelation function, which is less straightforward but much faster. The systematic best fit procedure is presented in a forthcoming paper, where we investigate also the three $\theta$'s case and show that it provides an excellent best fit of DNS spectra of the highest existing resolution (that is, the spectra of the higher moments are encapsulated with impressive accuracy).

In the present work we shall take a shortcut, which turns out to work rather well. We take the exponents to be equal, $\zeta_1=\zeta_2=1/3$, clearly assuming also the 2/3 law, so that there are two free parameters to be determined. Then we fiddle by hand the value of $C_2$ and $S$ in their expected intervals: $C_2$ is looked for in the neighborhood and above the value of 2, while $S$ is looked for in the neighborhood of the value given by the DNS data (Table 1) for each case, with the aim to bring the generalized Batchelor spectrum (13) as close as it is visibly possible to the respective DNS spectrum. This is done for the compensated spectrum $k^{5/3}E(k)$, in log-linear graphs. This is in accordance with our point of view regarding the nature of the Batchelor function: It is a model of the universal equilibrium range autocorrelation function, therefore the minimal requirement on it is to be able to approximate well the compensated (bottleneck) spectrum $k^{5/3}E(k)$, which gives also a good dissipation spectrum $k^2E(k)$ as a bonus. This essentially amounts to zooming at the energy spectrum $E(k)$ in a particular interval of the wavenumbers; in particular, the left-most interval of the wavenumbers in the universal equilibrium range. Then, by adding more small scales $\theta$ one may proceed to improve the approximation of the energy spectrum $E(k)$ deeper in the dissipation sub-range. Of course, when one works with a systematic best fit procedure what we described here arises automatically, and it is visible when looking in sequence the bottleneck spectrum $k^{5/3}E(k)$, the dissipation spectrum $k^2E(k)$, the palinstrophy spectrum $k^4E(k)$ and so on, in log-linear graphs. Presumably, the log-linear graphs emphasize naturally the interval of wavenumbers which is most strongly associated with each particular moment spectrum.

The results of our endeavors, first in terms of numbers, are shown in Table 1. We consider four cases with quoted Taylor-Reynolds numbers in the range of 250 to 1100. The value of $C_2$ ranges from 2.00 at the highest Reynolds number to 2.15 for the lowest Reynolds number. This is a reasonable range as well as pattern. The inertial range is less and less well formed as we look at lower Reynolds numbers, and lies effectively higher in a compensated spectrum graph because it is drawn upwards by the bottleneck bump. The bottleneck is a strong characteristic and in some sense precedes the inertial sub-range: When we look at the compensated spectrum at low Reynolds numbers all we see is the bottleneck bump. This phenomenon, along with the forcing on the flow, obstructs the formation of a uniform inertial range. More importantly, whatever it is formed depends on the Reynolds number. Therefore the value of $C_2$ (or equivalently, the value of $C_K$) is a parameter running with the Reynolds number. The values of the skewness $S$ were tuned by our best-fit-by-inspection procedure very close to the values quoted in reference [20]. The mismatch observed is due to three factors. One reason is the inadequacies of the best fit procedure. Second, the order of generalized Batchelor function: as explained above, two $\theta$'s cannot encapsulate the palinstrophy spectrum $k^4E(k)$, therefore the values of palinstrophy and skewness produced by the best fit are expected to be somewhat off the correct value in general. Third reason is the relatively low resolution of the DNS data, which means that the quoted values of the palinstrophy and skewness are affected by not negligible errors. In fact, the estimated and the quoted values for skewness could easily differ by 10 percent, as there is a difference of that order between the values for skewness from the DNS for $k_{max}\eta \approx 1$ and $k_{max}\eta \approx 2$ resolution level [20], that is, an error of that order in the DNS values of $S$ quoted below. It is surprising – although most possibly incidental – that the overall relative differences are much smaller.

| DNS data | | Model data | | | |
|---|---|---|---|---|---|
| $Re_\lambda$ | $S$ | $C_2$ | $S$ | $\theta_1/\eta$ | $\theta_2/\eta$ |
| 257 | 0.52 | 2.15 | 0.52 | 22.25 | 8.230 |
| 471 | 0.56 | 2.09 | 0.54 | 21.72 | 8.082 |
| 732 | 0.58 | 2.05 | 0.57 | 21.76 | 7.836 |
| 1131 | 0.60 | 2.00 | 0.61 | 21.78 | 7.542 |

**Table 1.** DNS data and Batchelor function input parameters

On the other hand, the bottleneck and dissipation spectra are excellent; they are shown in Figure 1 and 2, respectively. In both graphs the continuous jagged lines are the DNS spectra, and the thick dotted lines are the corresponding generalized Batchelor function spectra, for the four cases listed in Table 1. The input parameters for the generalized function, that is, the pair of $\theta/\eta$ or equivalently the value of $C_2$ and $S$, are quoted in the Table. The differences between the DNS and the generalized Batchelor spectra arise for wavenumbers on the left of the bottleneck peak where the DNS spectra exhibit their attempt to form an inertial sub-range (clearly shown in Figure 1) in the presence of forcing in a periodic box, and on the right-most part of the curves (in both Figures 1 and 2) i.e. the part of the spectrum that determines the shape of the palinstrophy spectrum $k^4 E(k)$. The palinstrophy spectra are shown in Figure 3. As emphasized just above, this is the part of the DNS spectrum which we cannot accommodate with the two $\theta$'s Batchelor function, while simultaneously we have reached the resolution limits of the DNS which can provide us only with an estimate of the *value* of the palinstrophy. We conclude that the two $\theta$'s of the generalized Batchelor function is an adequate order of approximation for turbulence at the $k_{max}\eta \approx 1$ resolution.

The thin dotted lines in Figures 1, 2 and 3 derive from the spectra associated to the usual (single $\theta$) Batchelor function, constructed with the same values for $C_2$ used for the generalized function (Table 1). This is especially clear in the graphs of Figure 1, where the left-most part of the thick and thin dotted curves coincide. In the language of the usual Batchelor function these reasonable values of $C_2$ translate to skewness 0.38 or less, as explained at the beginning of this section. This means that the palinstrophy is also small, causing the steeper downslope of the thin dotted curves in the Figures 1 and 2, and the smaller area under the curves of Figure 3. If, instead, we tried to tune skewness better, then two things would happen. First, the inertial sub-range part of the curves in Figure 1 would be too low: in the range of values for the skewness given in Table 1, $C_2$ would be smaller than 1.6 which corresponds to $C_K$ smaller than 1.2. Secondly, the bottleneck would be off position and at a different height. One could of course attempt a compromise between the two limits, or worse, attempt to zoom deeper in the large wave-numbers, trying to best fit the dissipation or the palinstrophy spectrum alone. But all that simply illustrates the fact that the single $\theta$ Batchelor function does not have enough small scale structure. We have shown that things improve considerably if we enrich that structure by one more scale, working with the two-factor form of the product expansion of (9). More specifically, this way we are able to capture the characteristics of the compensated spectrum $k^{5/3}E(k)$ of an actual spectrum, and then start progressing to model that spectrum deeper towards the dissipation sub-range. That is, the energy spectrum of the Batchelor function is let to live in its natural habitat i.e. the universal equilibrium range as a whole.

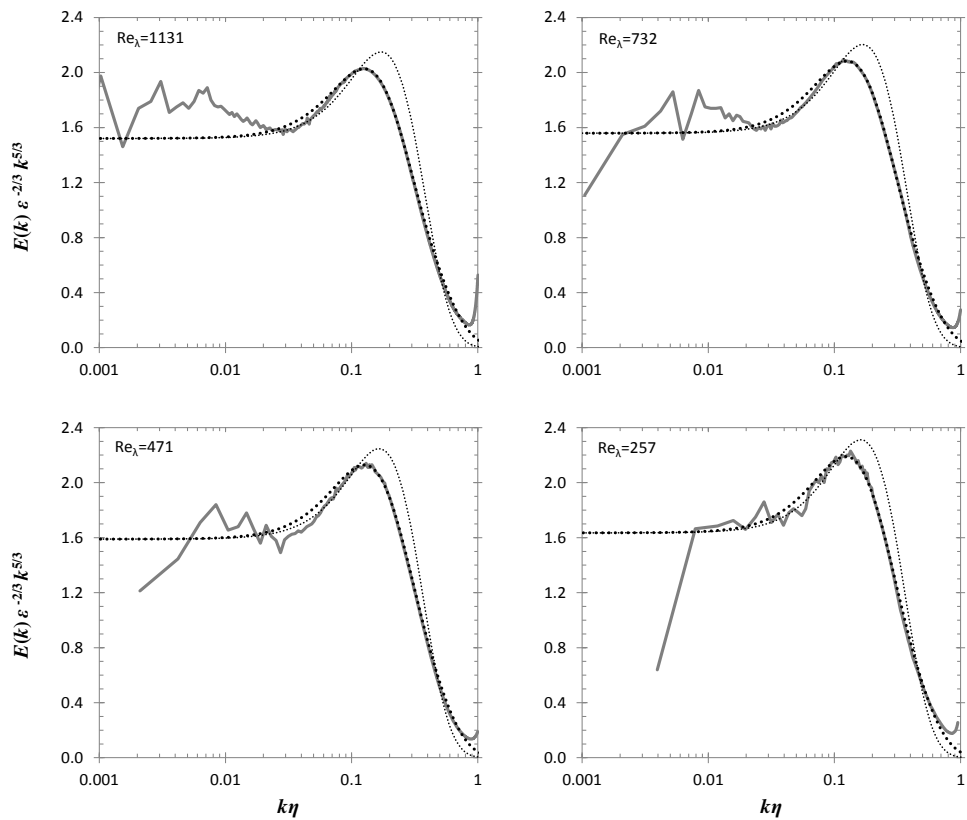

**Figure 1.** Compensated spectrum curves; continuous gray line: DNS data, thick dotted line: two $\theta$'s Batchelor spectrum, thin dotted line: usual (single $\theta$) Batchelor spectrum.

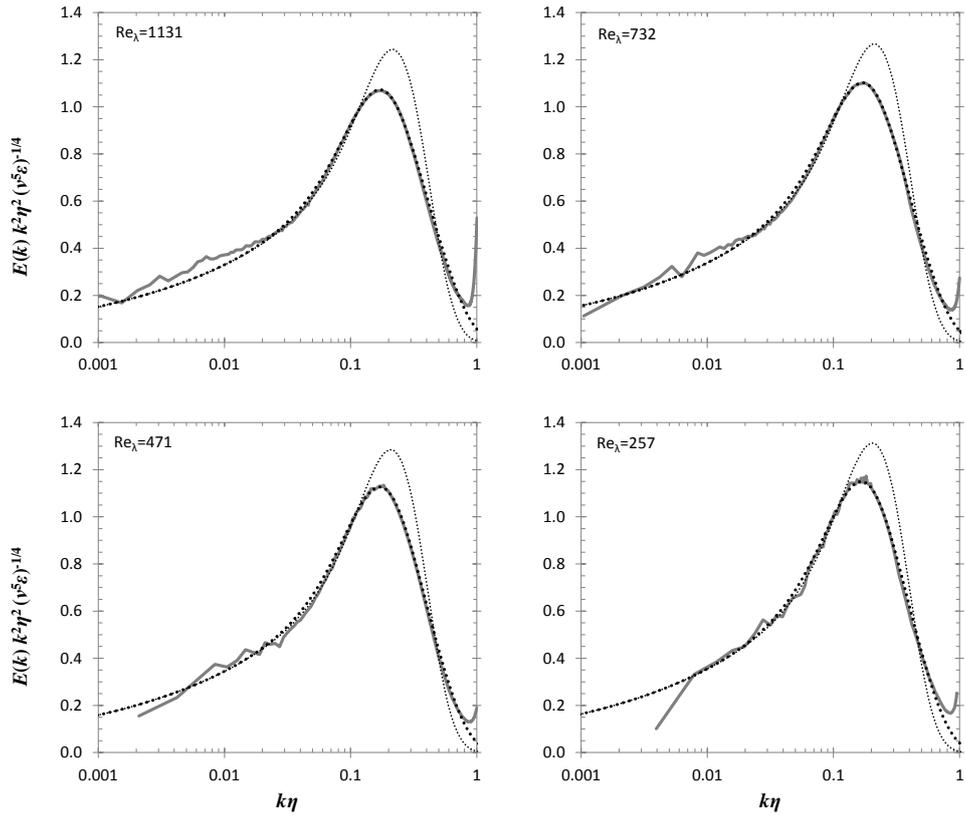

**Figure 2.** Dissipation spectrum curves; continuous gray line: DNS data, thick dotted line: two $\theta$'s Batchelor spectrum, thin dotted line: usual (single $\theta$) Batchelor spectrum.

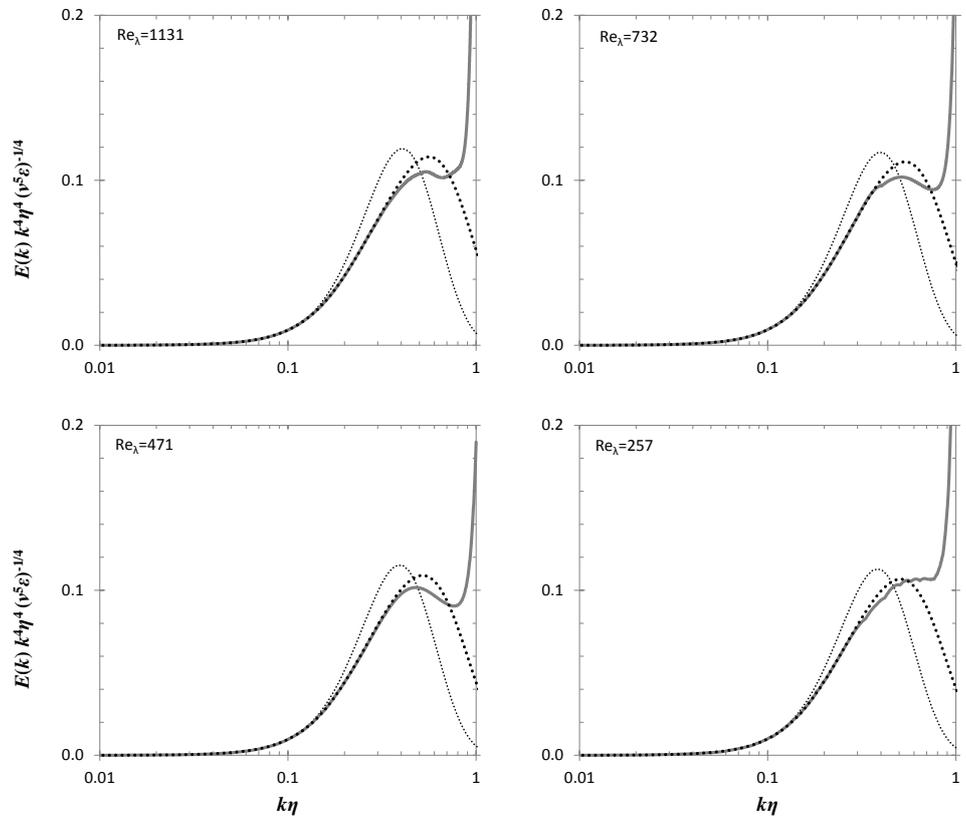

**Figure 3.** Palinstrophy spectrum curves; continuous gray line: DNS data, thick dotted line: two $\theta$'s Batchelor spectrum, thin dotted line: usual (single $\theta$) Batchelor spectrum.

## IV. DISCUSSION

The generalized Batchelor function given by equation (9) is proposed as a model for the longitudinal autocorrelation function of isotropic turbulence in the universal equilibrium range. The degree of accuracy of the model is controlled by the number of factors in the product expansion of (9) i.e. in the number of small length scales $\theta$. The conjecture is that a potentially infinite product could amount to the 'exact' result. In practice, and what matters most, is that the number of $\theta$'s is related to the number of moments of a given actual spectrum which are known with sufficient accuracy. That is, it may be said somewhat loosely, the spectrum of the generalized Batchelor function is accurate to the degree the actual spectrum which is modelled is known accurately. [Interestingly, one may note that there is an affinity between this behavior and characteristics of the renormalization group approach to turbulence [16]: the fixed resolution level encoded in a (generalized) Batchelor spectrum is an analogue of the cut-off scale in that approach, while the association of the resolution level with the number of $\theta$ and the value of the higher moments is an analogue and the running scale dependence of the physical quantities in the same approach.] We applied these ideas to high-resolution direct numerical simulations (DNS) of turbulence [19][20][21] which include the highest Reynolds number DNS flows achieved. We used the two $\theta$'s case of the generalized function, in consistency with the degree of accuracy of the DNS spectra, for which the only higher moment determined with decent accuracy is palinstrophy. The result is that the compensated (bottleneck) spectra $k^{5/3}E(k)$ and the dissipation spectra $k^2E(k)$ of the DNS are captured very well, while the palinstrophy DNS spectrum $k^4E(k)$ may only be partially captured, as it is ill determined due to the DNS resolution limits. Specific application of generalized Batchelor functions with larger product expansions, such as that of three $\theta$'s, is the subject of work under preparation. Using DNS data of large Reynolds numbers less than the maximum available one may test the conjecture that at the three $\theta$'s product expansion the palinstrophy spectrum will be captured very well. In fact, preliminary investigations show that even the next higher moment spectrum is modelled very nicely. This supports our point of view that the Batchelor function, in its full proposed form, is sufficient to model isotropic turbulent flows in their universal equilibrium range, modulo effects related to the setup of the flows, such as forcing.